\documentclass[conference]{IEEEtran}
\IEEEoverridecommandlockouts

\usepackage{cite}
\usepackage[colorlinks=true,
            linkcolor=blue,
            citecolor=blue,
            urlcolor=blue]{hyperref}
            
\usepackage{amsmath,amssymb,amsfonts}
\usepackage{algorithmic}
\usepackage{graphicx}
\usepackage{textcomp}
\usepackage{xcolor}

\usepackage{braket}
\usepackage{booktabs}
\usepackage{listings}

\def\BibTeX{{\rm B\kern-.05em{\sc i\kern-.025em b}\kern-.08em
    T\kern-.1667em\lower.7ex\hbox{E}\kern-.125emX}}

\title{Physics-Informed Discrete-Event Simulation of Polarization-Encoded Quantum Networks
\thanks{Any mention of commercial products or references to commercial organizations is for information only; it does not imply
recommendation or endorsement by NIST, nor does it imply that the products mentioned are necessarily the best available for the purpose}
}

\author{
\centering
    \IEEEauthorblockN{
        Abderrahim Amlou\IEEEauthorrefmark{1}\IEEEauthorrefmark{2},
        Amar Abane\IEEEauthorrefmark{1},
        Cory M. Nunn\IEEEauthorrefmark{1},
        M. V. Jabir\IEEEauthorrefmark{1},
        Van Sy Mai\IEEEauthorrefmark{1},
        Abdella Battou\IEEEauthorrefmark{1},
        Ahmed Lbath\IEEEauthorrefmark{2}
    }
    \IEEEauthorblockA{
        \IEEEauthorrefmark{1}National Institute of Standards and Technology, Gaithersburg, MD, USA\quad
        \IEEEauthorrefmark{2}Univ. Grenoble Alpes, Grenoble, France\\
            abderrahim.amlou@nist.gov, amar.abane@nist.gov, cory.nunn@nist.gov, jabir.marakkarakathvadakkepurayil@nist.gov,\\
            vansy.mai@nist.gov, abdella.battou@nist.gov,
            ahmed.lbath@univ-grenoble-alpes.fr
    }
}

\begin{document}

\maketitle

\begin{abstract}
We extend the SeQUeNCe discrete-event simulator with physics-based models for polarization-encoded photonic quantum networks. Our framework integrates Jones-calculus optical components, including an SPDC Bell-state source, wave plates, and polarizing beam splitters, together with a multi-section fiber model capturing polarization mode dispersion, chromatic dispersion, and Raman noise from coexisting classical traffic. We validate the simulator by reproducing experimentally reported spectra, polarization correlations, quantum state tomography, and dispersion- and Raman-induced noise. The resulting platform enables hardware-parameterized prediction of entanglement distribution performance under realistic deployment conditions.
\end{abstract}


\section{Introduction}
The realization of large-scale quantum networks capable of distributing high-fidelity entanglement over metropolitan and intercity distances represents a foundational milestone for quantum communication~\cite{wengerowsky2019entanglement,zhang2008distribution}. 
Recent experiments demonstrated polarization entanglement distribution over 96~km of submarine fiber~\cite{wengerowsky2019entanglement}, time-bin entanglement over 100~km of fiber spools~\cite{zhang2008distribution}, four-dimensional entanglement over 100~km~\cite{ikuta2018four}, and polarization entanglement with classical-quantum coexistence over 100~km of deployed metropolitan fiber~\cite{rahmouni2024100,burenkov2023synchronization}. 
These advances bring practical quantum networks closer to reality. 

Translating these demonstrations into robust architectures requires discrete-event simulators such as SeQUeNCe~\cite{wu2021sequence}, NetSquid~\cite{coopmans2021netsquid}, and Multiverse~\cite{abane2025multiverse}, which model network topologies, routing protocols, and entanglement distribution at scale.
Nevertheless, optical fiber links are typically modeled as delay elements with fixed attenuation and phenomenological depolarization parameters. While suitable for protocol-level evaluation, such abstractions obscure the dominant physical processes governing real polarization-encoded quantum links. 
In deployed fibers, polarization evolves continuously under the influence of stochastic birefringence, temperature fluctuations, bending-induced stress, and polarization mode dispersion (PMD). In metropolitan deployments, co-propagating classical channels introduce wavelength-dependent Raman-scattering noise whose statistics depend on launch power, spectral separation, and fiber length~\cite{burenkov2023synchronization,rahmouni2024100}. 
These effects are not merely implementation details: they dynamically alter entanglement fidelity, coincidence rates, and the stability of polarization correlations, which are typically extracted from time-tagged detection data in deployed entanglement-distribution experiments~\cite{amlou2025scalable}.

The absence of physics-based optical modeling in network simulators creates a cross-layer blind spot, particularly for polarization-encoded quantum networks. Although polarization is a natural and widely adopted encoding for photonic qubits~\cite{bennett1992quantum,ekert1991quantum}, its fidelity over deployed fiber is highly sensitive to stochastic birefringence, polarization mode dispersion (PMD), and environmental perturbations such as temperature and mechanical stress~\cite{hubel2007high}. 
Recent advances in first-principles modeling of fiber birefringence, such as the BIFROST framework~\cite{banner2025bifrost}, have linked polarization evolution to underlying material and geometric parameters. Other experimental studies have quantified Raman-scattered noise as a function of classical wavelength and fiber length.~\cite{burenkov2023synchronization,rahmouni2024100}. 
However, these developments remain disconnected from discrete-event quantum network simulation frameworks.

In this work, we close this gap by integrating physics-based polarization and fiber-channel modeling into the SeQUeNCe simulator \footnote{Code available upon request.}. We replace abstract depolarization models with Jones-matrix-based optical components and stochastic, temperature and geometry-dependent fiber models. Our framework captures polarization evolution, chromatic dispersion, PMD, and Raman scattering within a unified discrete-event architecture parameterized by experimentally measurable quantities.

Our contributions are threefold. First, we introduce a modular library of polarization-resolved optical components, enabling construction of realistic entanglement-distribution experiments within SeQUeNCe. Second, we implement a comprehensive fiber model that combines temperature-dependent Sellmeier-based dispersion, stochastic birefringence and PMD arising from geometry and stress, and wavelength-dependent Raman noise induced by co-propagating classical channels. Third, we validate our implementation by reproducing experimentally observed SPDC spectra, polarization-correlation fringes, quantum state tomography, PMD-induced temporal broadening, and Raman-noise trends reported in~\cite{farella2024spectral,burenkov2023synchronization}.

By embedding optical-layer physics into SeQUeNCe, our work establishes a cross-layer modeling framework that connects fiber-level physical parameters to network-level performance metrics. 
This enables systematic evaluation of how deployment conditions, environmental variability, and classical coexistence constraints impact entanglement fidelity, link stability, and protocol behavior in realistic metropolitan quantum networks.

The remainder of this paper is organized as follows. Section~\ref{sec:background} reviews the relevant polarization and fiber physics and summarizes the SeQUeNCe framework. Sections~\ref{sec:components} and~\ref{sec:fiber} introduce our quantum optical components and physics-based fiber model, respectively. Section~\ref{sec:validation} presents validation against experimental results. Section~\ref{sec:discussion} discusses limitations and future directions, and Section~\ref{sec:conclusion} concludes the paper.

\section{Background}
\label{sec:background}
In this section we briefly review the mathematical and physical foundations underlying fiber-optic quantum channels and the aspects of the SeQUeNCe simulator that are relevant for our extensions.

\subsection{Jones calculus and polarization optics}
We describe single-photon polarization states using the Jones formalism. In a fixed transverse basis of horizontal and vertical polarization, the state of a photon is written as a two-component column complex vector
\begin{equation}
\ket{\Psi} =
\begin{pmatrix}
\alpha_H \\
\alpha_V
\end{pmatrix},
\label{eq:jones_state}
\end{equation}
with \(|\alpha_H|^2 + |\alpha_V|^2 = 1\), where \(|\alpha_H|^2\) and \(|\alpha_V|^2\) are the detection probabilities \cite{jones1941new,le1997utilization} respectively for\footnote{Other common polarization states, such as diagonal, anti-diagonal, right- and left-circular, are simple superpositions of \(\ket{H}\) and \(\ket{V}\).}: 

\[
\ket{H} =
\begin{pmatrix}
1 \\
0
\end{pmatrix},
\qquad
\ket{V} =
\begin{pmatrix}
0 \\
1
\end{pmatrix}.
\]

Polarization transformations induced by lossless optical elements are represented by unitary Jones operators \(\mathbf{J}\in \mathrm{U}(2)\) acting on the single-photon polarization state $\ket{\Psi_{\text{out}}} = \mathbf{J}\,\ket{\Psi_{\text{in}}}$ with cascaded elements described by operator composition. Specific physical effects correspond to particular realizations of \(\mathbf{J}\).
When acting on a two-photon polarization state, the transformation is lifted to the composite space as $\mathbf{J}\otimes\mathbf{I}$ or $\mathbf{I}\otimes\mathbf{J}$, depending on the photon addressed.

\subsection{Polarization Effects in Fiber}
Commercial single-mode telecom fibers are designed to be cylindrically symmetric, but in practice they have a small residual birefringence from manufacturing imperfections and environmental perturbations. As light propagates, the two orthogonal polarization components see slightly different refractive indices, so one effectively travels a bit faster than the other. Their relative phase and arrival time therefore drift along the fiber, a phenomenon known as polarization mode dispersion (PMD), which leads to polarization rotation and differential group delay (DGD) that grow with distance and depend on temperature, mechanical stress, and wavelength. 

As a result, each photon experiences polarization transformation dependent on its wavelength and position, and its propagation speed can change between fiber sections~\cite{gordon2000pmd}.
These effects can be related quantitatively to underlying physical parameters, such as fiber core ellipticity, material composition and mechanical stress using first-principles models like BIFROST~\cite{banner2025bifrost}. 


\subsection{Chromatic Dispersion in Fiber}

Chromatic dispersion arises from the wavelength dependence of the refractive index of glass, causing different wavelengths to propagate at different group velocities. The refractive index of optical materials is commonly modeled using the Sellmeier equation \cite{banner2025bifrost}, which relates the index to wavelength through a series of resonances:
\begin{equation}   
n^2(\lambda) = 1 + \sum_{i=1}^{3} \frac{B_i \lambda^2}{\lambda^2 - C_i^2}
\label{eq:sellmeier}
\end{equation}
where \(B_i\) (unitless) are resonance strengths and \(C_i\) (in meters) are resonance wavelengths. For silica glass, these resonances correspond to UV electronic transitions and infrared vibrational modes. Temperature-dependent formulations extend this model by making \(B_i\) and \(C_i\) functions of temperature \cite{banner2025bifrost}.

The chromatic dispersion of a fiber is quantified by the dispersion parameter \(D_{\text{CD}}\) (in ps/(nm\(\cdot\)km)), which can be approximated by the empirical formula \cite{banner2025bifrost}:
\begin{equation}   
D_{\text{CD}}(\lambda) = \frac{S_0}{4}\left[\lambda - \frac{\lambda_0^4}{\lambda^3}\right]
\label{eq:chromatic_dispersion_formula}
\end{equation}
where \(\lambda_0\) is the zero-dispersion wavelength and \(S_0\) is the dispersion slope at \(\lambda_0\). This formula combines both material and waveguide dispersion effects and is widely used in fiber specifications. Then the chromatic dispersion delay relative to the fiber's reference wavelength $\lambda_{\text{ref}}$ is
\begin{equation}    
\tau_{\text{CD}} = D_{\text{CD}}(\lambda_{\text{ref}}) \cdot L \cdot (\lambda_q - \lambda_{\text{ref}})
\label{eq:chromatic_dispersion_delay}
\end{equation}
where $D_{\text{CD}}(\lambda_{\text{ref}})$ is the dispersion parameter evaluated at the reference wavelength.



\subsection{Raman scattering and classical–quantum coexistence}
In many deployments, quantum links share the same fiber as classical channels used for synchronization, control, or data traffic, because this greatly simplifies synchronization, deployment and reduces cost.~\cite{rahmouni2024100}. High-power classical signals then act as a source of broadband noise through spontaneous Raman scattering in the fiber~\cite{agrawal2000nonlinear}. Raman scattering redistributes light from the classical channel into a broad range of other wavelengths, both longer and shorter than the original wavelength, and some of these scattered photons can fall inside the quantum receiver’s passband.

Experiments have characterized the resulting noise in terms of effective forward- and backward-scattering coefficients that depend on the classical and quantum wavelengths and on the fiber type~\cite{burenkov2023synchronization}. For a fixed quantum channel near 1550~nm, these measurements show that the noise can vary by almost two orders of magnitude between different classical wavelengths (for example, 1270~nm versus 1490~nm) for the same launched power and length~\cite{burenkov2023synchronization,rahmouni2024100}. These results indicate that coexistence models in simulation must explicitly account for the dependence of Raman noise on both wavelength and fiber length, rather than treating the classical background as a simple, fixed dark-count term.

\subsection{SeQUeNCe simulator}
SeQUeNCe is an open-source discrete-event simulator for quantum networks~\cite{wu2021sequence}. It provides a timeline-driven simulation engine, node and channel abstractions, and a quantum state manager that supports several encodings, including polarization. Quantum channels in the base package model fiber as a link with fixed attenuation, propagation delay, and an optional abstract depolarization probability that does not depend on wavelength, temperature, or classical traffic.

This abstraction is adequate for many protocol-level studies, but it cannot represent polarization evolution, PMD, or Raman-scattering noise in a physically meaningful way. In this work we retain SeQUeNCe’s event-driven architecture and node/channel interfaces but replace the abstract polarization treatment with a Jones-matrix-based fiber model and realistic component models tailored to polarization-encoded entanglement distribution over metropolitan distances.

\section{Quantum Optical Components}
\label{sec:components}
In this section, we describe the implemented components on top of SeQUeNCe to build end-to-end simulations of polarization-encoded entanglement distribution over fiber.

\subsection{SPDC Bell-state source}
\label{sec:spdc_source}
Spontaneous parametric down-conversion (SPDC) in nonlinear crystals is a standard way to generate entangled photon pairs for quantum communication~\cite{kwiat1995new}. In type-II SPDC, a pump photon is converted into a signal–idler pair with orthogonal polarizations and frequencies \(\omega_s\) and \(\omega_i\) that satisfy energy conservation \(\omega_p = \omega_s + \omega_i\). By appropriate crystal and interferometer settings, this process can produce any of the four maximally entangled Bell states defined in the two-photon polarization basis $\{\ket{HH}, \ket{HV}, \ket{VH}, \ket{VV}\}$ ~\cite{kwiat1995new,james2001measurement}.

Our \texttt{SPDCBellSource} class extends SeQUeNCe's existing \texttt{LightSource} to model a pulsed SPDC source that emits polarization-entangled pairs. The main additions are explicit Bell-state generation, configurable photon-pair statistics (thermal or Poisson), and wavelength correlations between signal and idler photons.

\subsubsection*{Bell states}
Each emitted pair is represented as a four-component state vector in the computational basis \(\{\ket{HH},\ket{HV},\ket{VH},\ket{VV}\}\). The source can be configured to produce any of the four standard Bell states:
\begin{align*}
\ket{\Phi^+} &= \tfrac{1}{\sqrt{2}}\bigl(\ket{HH} + \ket{VV}\bigr)
&\quad
\ket{\Phi^-} &= \tfrac{1}{\sqrt{2}}\bigl(\ket{HH} - \ket{VV}\bigr) \\
\ket{\Psi^+} &= \tfrac{1}{\sqrt{2}}\bigl(\ket{HV} + \ket{VH}\bigr)
&\quad
\ket{\Psi^-} &= \tfrac{1}{\sqrt{2}}\bigl(\ket{HV} - \ket{VH}\bigr)
\end{align*}
with the choice set by a configuration parameter. When a pair is emitted, two photon objects (labeled signal and idler) are created and registered as entangled in the quantum state manager.

\subsubsection*{Photon-pair statistics}
The number of photon pairs produced per pump pulse depends on the pumping regime. For a coherent pump laser, joint coincidence measurements typically follow Poissonian statistics~\cite{carvalho2025spectrally},
\[
P_{\text{Pois}}(n) = \frac{\mu^n e^{-\mu}}{n!},
\]
where \(n\) is the number of pairs and \(\mu\) is the mean. However, when measuring the signal (or idler) beam alone with spectral filtering and short measurement windows, thermal (Bose–Einstein) statistics can emerge~\cite{carvalho2025spectrally},
\[
P_{\text{th}}(n) = \frac{1}{1+\mu}\left(\frac{\mu}{1+\mu}\right)^n,
\]
which better captures the correlated, bunched nature of SPDC emission under these conditions. SeQUeNCe's base source implements Poissonian statistics; we add thermal statistics to model this experimentally observed behavior.

\subsubsection*{Wavelength correlation}
In the SPDC source model, the signal wavelength $\lambda_s$ is sampled from a truncated Gaussian distribution centered at a user-defined mean $\lambda_{s,0}$, with a standard deviation set by the configured bandwidth. The idler wavelength $\lambda_i$ is then deterministically computed to satisfy energy conservation,
\begin{equation}
\frac{1}{\lambda_p} = \frac{1}{\lambda_s} + \frac{1}{\lambda_i},
\end{equation}
where $\lambda_p$ is the pump wavelength. These correlated wavelengths are later used by the fiber model to compute chromatic dispersion for each photon independently.

\subsubsection*{Interface}
The source inherits the timing mechanism from \texttt{LightSource}, with emission pulses spaced by the inverse of the configured repetition rate. The Bell state, mean pair number, photon statistics, and wavelength pair can all be changed at runtime, so protocol studies that sweep source brightness or switch between different entangled states do not require rebuilding the network model.

\subsection{Wave plates}
\label{sec:wave_plates}
Wave plates are birefringent optical elements that manipulate photon polarization through a controlled phase retardation between orthogonal components~\cite{jones1941new}. We implement half-wave plates (HWP) and quarter-wave plates (QWP) using Jones matrix formalism to manipulate single-photon and entangled two-photon polarization states.

\subsubsection*{Jones matrix representation}
For a wave plate rotated by angle \(\theta\) relative to the horizontal axis, the transformation is described by a \(2 \times 2\) Jones matrix acting on the polarization state vector \(\ket{\Psi} = \alpha\ket{H} + \beta\ket{V}\). The matrices are:

\textbf{Half-wave plate:}
\begin{equation}
\mathbf{J}_{\text{HWP}}(\theta) = \begin{pmatrix} \cos 2\theta & \sin 2\theta \\ \sin 2\theta & -\cos 2\theta \end{pmatrix}
\label{eq:hwp}
\end{equation}

\textbf{Quarter-wave plate:}
\begin{equation}
\mathbf{J}_{\text{QWP}}(\theta) = \begin{pmatrix} \cos^2\theta + i\sin^2\theta & (1-i)\cos\theta\sin\theta \\ (1-i)\cos\theta\sin\theta & \sin^2\theta + i\cos^2\theta \end{pmatrix}
\label{eq:qwp}
\end{equation}

A half-wave plate rotates linear polarization by \(2\theta\) without changing the polarization type, while a quarter-wave plate converts between linear and circular/elliptical polarization states depending on \(\theta\).


\subsubsection*{Interface}
The \texttt{WavePlate} component accepts a single input and forwards transformed photons to a single output. The plate type (HWP or QWP), rotation angle \(\theta\), and fidelity can be specified at initialization or updated dynamically during simulation, which enables active polarization control and reconfigurable optical circuits.

\subsection{Polarizing beam splitter}
\label{sec:pbs}
A polarizing beam splitter (PBS) routes photons to different output ports based on their polarization state. In quantum communication systems, the PBS acts as a non-demolishing polarization measurement device that preserves photon number while determining polarization~\cite{kwiat1995new}. We implemented a fixed-orientation PBS with realistic error modeling.

\subsubsection*{Fixed-basis measurement}
The PBS is configured with a measurement basis index that determines its orientation. For basis index 0, the PBS measures in the rectilinear $\{\ket{H}, \ket{V}\}$ basis; for basis index 1, it measures in the diagonal $\{\ket{+}, \ket{-}\}$ basis.

When a photon arrives, the PBS performs a projective measurement in its configured basis. The measurement outcome (0 or 1) determines which of two output ports receives the photon.  In the absence of errors, the operation is
\begin{equation}
\ket{\Psi} \xrightarrow{\text{PBS}} 
\begin{cases}
\text{port 0} & \text{with probability } |\langle b_0 | \Psi \rangle|^2 \\
\text{port 1} & \text{with probability } |\langle b_1 | \Psi \rangle|^2
\end{cases}
\label{eq:pbs_measurement}
\end{equation}
where $|b_0\rangle$ and $|b_1\rangle$ are the two orthogonal states defining the measurement basis: $\{\ket{H}, \ket{V}\}$ for basis index 0, or $\{\ket{+}, \ket{-}\}$ for basis index 1.

\subsubsection*{Error modeling}
Real PBS components suffer from imperfect polarization discrimination, which leads to measurement errors. Commercial PBS cubes typically achieve extinction ratios of 100:1 to 10,000:1, meaning the unwanted polarization component is suppressed by this factor relative to the desired component~\cite{thorlabs_pbs}. Rather than model the full extinction ratio, we introduce a bit-flip error probability $p_{\text{err}}$ that captures the net effect on protocol performance: after the ideal projective measurement, the outcome is flipped with probability $p_{\text{err}}$, which routes the photon to the wrong output port.

The PBS also includes SeQUeNCe's standard fidelity parameter $f \in [0,1]$ to model insertion loss, where each photon is successfully transmitted with probability $f$ and lost with probability $1-f$.

\subsubsection*{Interface}
The \texttt{PolarizingBeamSplitter} accepts a single input and routes photons to one of two outputs based on the measurement result. The measurement basis, transmission fidelity, and measurement error probability are specified at initialization and remain constant throughout the simulation. For complete polarization measurement, the PBS is combined with two single-photon detectors (already provided by SeQUeNCe) in the \texttt{QSDetectorPolarizationStatic} class, which provides the interface for static-basis Bell state measurements used in entanglement distribution protocols.

\subsection{High-level node abstractions}
\label{sec:node_abstractions}
\begin{figure*}[t]
    \centering
    \includegraphics[width=0.8\linewidth]{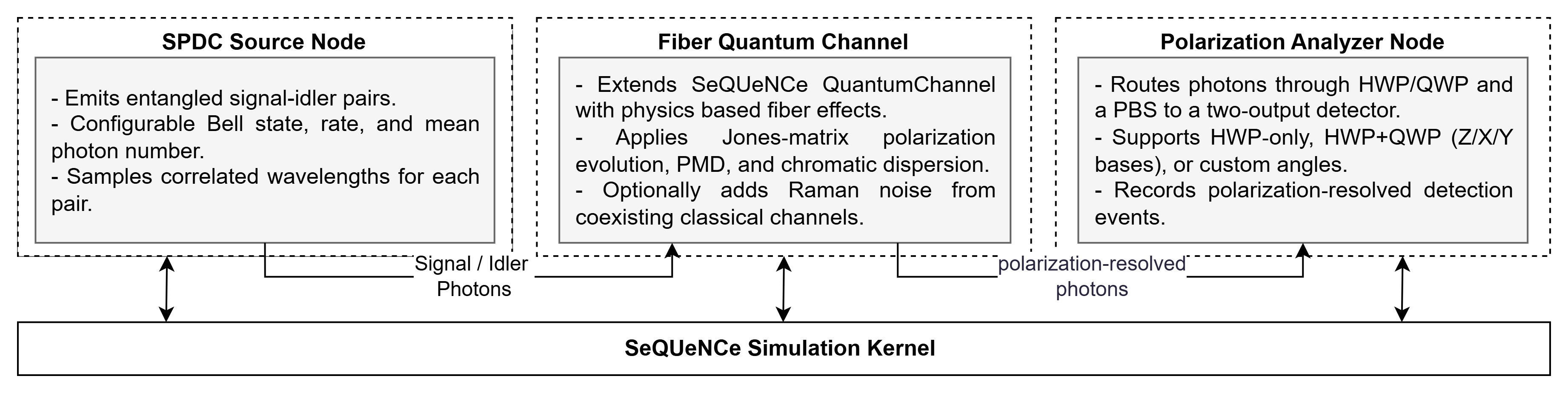}
    \caption{Integration of the optical components with the SeQUeNCe simulator.}
    \label{fig:architecture}
\end{figure*}

To simplify simulation setup, we provide composite node classes that encapsulate common experimental configurations. Figure \ref{fig:architecture} summarizes the new optical components and their integration with the SeQUeNCe kernel.

The \texttt{Spdc Source Node} wraps the SPDC source with network routing infrastructure and exposes methods like \texttt{set\_bell\_state()} and \texttt{set\_mean\_photon\_num()} for runtime parameter control.

The \texttt{Polarization Analyzer Node} combines wave plates and the static-basis detector into three standard measurement architectures. In \emph{HWP-only mode}, a single half-wave plate enables linear polarization rotation for rectilinear/diagonal basis switching. In \emph{HWP+QWP mode}, the combination of both wave plates enables full polarization tomography, with a \texttt{set\_basis()} method that automatically configures angles to measure in the Pauli \(Z\), \(X\), or \(Y\) bases. A \emph{custom mode} allows arbitrary wave plate configurations for non-standard measurement schemes. All hardware parameters (efficiency, dark counts, insertion loss, measurement errors) are configurable through a single dictionary interface at node creation.
The physics‑based fiber quantum channel is detailed in Section ~\ref{sec:fiber}.

\section{Physics-Based Fiber Channel Model}
\label{sec:fiber}
To accurately simulate photon propagation in optical fibers, we implemented a comprehensive fiber channel model that accounts for polarization evolution, temporal dispersion, and optical noise. Rather than using simplified phenomenological models, we adopted the physics-based approach described in recent literature \cite{banner2025bifrost,burenkov2023synchronization}, which grounds fiber behavior in measurable material properties and environmental parameters.

\subsection{Model Overview and Rationale}
Our fiber channel model incorporates three primary physical effects: (1) polarization mode dispersion due to fiber birefringence, (2) chromatic dispersion causing wavelength-dependent delays, and (3) Raman scattering noise when classical synchronization signals coexist with quantum channels. The model uses Jones matrix formalism (Section~\ref{sec:background}) to track polarization evolution and temperature-dependent Sellmeier equations (Eq.~\ref{eq:sellmeier}) for material properties, following the implementations validated in \cite{banner2025bifrost,burenkov2023synchronization}. We extend these single-fiber analytical models to support multi-section configurations with heterogeneous properties and integrate them into a discrete-event quantum network simulator.

Material refractive indices are computed using the temperature-dependent Sellmeier formulation and parameter values reported in \cite{banner2025bifrost}. Standard single-mode fiber parameters are used unless otherwise specified.



\subsection{Birefringence and Polarization Evolution}
Fiber birefringence is modeled as the combination of linear and circular contributions. These effects are defined \textit{locally per fiber section} and later composed to obtain the full fiber response.

The linear term represents birefringence induced by core ellipticity, thermal stress, and bending. Based on \cite{banner2025bifrost}, for each fiber section \(i\), the total linear birefringence is expressed as:
\begin{equation}
\Delta\beta_{\text{linear},i} =
\Delta\beta_{\text{ellip},i} +
\Delta\beta_{\text{thermal},i} +
\Delta\beta_{\text{bend},i}
\end{equation}
where $\Delta\beta_{\text{ellip},i}$ arises from core ellipticity due to manufacturing imperfections, $\Delta\beta_{\text{thermal},i}$ is induced by temperature-dependent stress, and $\Delta\beta_{\text{bend},i}$ results from mechanical bending of section \(i\)~\cite{banner2025bifrost}. 

Circular birefringence is similarly defined per section and arises from fiber twisting, denoted $\Delta\beta_{\text{twist},i}$.

The birefringence contributions are represented as Jones matrices for polarization tracking. Linear birefringence applies a diagonal phase transformation within each section.

For a fiber section \(i\) of length \(L_i\), the linear birefringence produces a phase retardation given by:
\begin{equation}
\mathbf{J}_{\text{lin},i} = \begin{pmatrix}
e^{i\Delta\beta_{\text{linear},i} L_i/2} & 0 \\
0 & e^{-i\Delta\beta_{\text{linear},i} L_i/2}
\end{pmatrix}
\label{eq:J_lin}
\end{equation}

Similarly, circular birefringence in section \(i\) produces a rotation matrix:
\begin{equation}
\mathbf{J}_{\text{circ},i} = \begin{pmatrix}
\cos(\Delta\beta_{\text{twist},i} L_i/2) & -\sin(\Delta\beta_{\text{twist},i} L_i/2) \\
\sin(\Delta\beta_{\text{twist},i} L_i/2) & \cos(\Delta\beta_{\text{twist},i} L_i/2)
\end{pmatrix}
\label{eq:J_circ}
\end{equation}

For each fiber section \(i\), the total Jones matrix is:
\begin{equation}
\mathbf{J}_{\text{section},i} = \mathbf{J}_{\text{circ},i} \cdot \mathbf{J}_{\text{lin},i}
\end{equation}
which combines both effects locally within the section~\cite{banner2025bifrost}.

After constructing the section-level matrices \(\mathbf{J}_{\text{section},i}\), the composite transformation for the entire fiber link is obtained by cascading all sections:
\begin{equation}
\mathbf{J}_{\text{total}}(\lambda) = \prod_{i=N}^{1} \mathbf{J}_{\text{section},i}(\lambda)
\end{equation}
where the ordered product is taken in reverse index order to match the photon propagation direction. 

Thus, birefringence is modeled as a sequence of local transformations, and \(\mathbf{J}_{\text{total}}\) represents the global polarization evolution across the full fiber.

This transformation is applied to each photon's polarization state as it enters the fiber.

\subsection{Differential Group Delay Calculation}
To quantify polarization mode dispersion (PMD), we compute the differential group delay (DGD) from the \textit{composite} Jones matrix \(\mathbf{J}_{\text{total}}(\lambda)\), i.e., the full end-to-end transformation of the fiber link.

DGD is obtained using an eigenvalue-based frequency perturbation method \cite{banner2025bifrost}. Specifically, we evaluate how the composite Jones matrix varies under a small spectral shift around the operating wavelength.

We compute the eigenvalues of:
\[
\mathbf{J}_{\text{total}}(\lambda \pm \delta\lambda)\mathbf{J}_{\text{total}}^{-1}(\lambda)
\]
where \(\delta\lambda\) is a small wavelength perturbation. This operation captures the relative phase evolution between the principal states of polarization.

The DGD is then estimated using a symmetric finite-difference approximation:
\begin{equation}
\text{DGD} = \frac{1}{2}\left[
\left|\frac{\arg(\mu_+^- / \mu_-^-)}{\delta\omega}\right| +
\left|\frac{\arg(\mu_+^+ / \mu_-^+)}{\delta\omega}\right|
\right]
\end{equation}

Here, \(\arg(\cdot)\) denotes the phase of the complex number, and the angular frequency is defined as \(\omega = 2\pi c / \lambda\). The perturbation is performed in wavelength, and the corresponding frequency step is obtained using a first-order (linear) approximation:
\[
\delta\omega \approx \left| \frac{d\omega}{d\lambda} \right| \delta\lambda = \frac{2\pi c}{\lambda^2} \delta\lambda
\]
In our implementation, we use \(\delta\lambda = 0.1\,\text{nm}\), which provides a good trade-off between numerical stability and accuracy. We verified that smaller step sizes yield negligible changes in the computed DGD, indicating convergence.

This approach estimates the differential group delay as the derivative of the phase difference between the two principal polarization modes with respect to frequency.

\subsection{Chromatic Dispersion}
Group velocity dispersion causes wavelength-dependent propagation delays. For each fiber section, we compute the dispersion parameter $D_{\text{CD},i}$ using the empirical formula from Eq.~\ref{eq:chromatic_dispersion_formula}. 

For multi-section fibers, the effective dispersion parameter for the entire link is computed as a length-weighted average:
\begin{equation}
D_{\text{CD,eff}} = \frac{\sum_{i=1}^{N} D_{\text{CD},i} \cdot L_i}{\sum_{i=1}^{N} L_i}
\end{equation}
where $D_{\text{CD},i}$ and $L_i$ are the dispersion parameter and length of section $i$, respectively. This weighted average correctly represents the accumulated chromatic dispersion of the composite link. The chromatic dispersion delay $\tau_{\text{CD}}$ for a photon at wavelength $\lambda_q$ is then computed once for the entire fiber using Eq.~\ref{eq:chromatic_dispersion_delay}:
\begin{equation}
\tau_{\text{CD}} = D_{\text{CD,eff}} \cdot L_{\text{total}} \cdot (\lambda_q - \lambda_{\text{ref}})
\end{equation}
where $L_{\text{total}} = \sum_{i=1}^{N} L_i$ is the total fiber length and $\lambda_{\text{ref}}$ is the reference wavelength (taken from the first section's specification). 

Each photon experiences a delay $\tau_{\text{CD}}$ that depends linearly on its wavelength deviation from the reference, with the total delay accumulated across all fiber sections.

\subsection{Raman Noise from Classical Coexistence}
When quantum channels share fiber with classical synchronization signals, inelastic Raman scattering generates background photons in the quantum channel. We implement the noise model from \cite{burenkov2023synchronization}, which characterizes both backward-scattered and forward-scattered contributions.

For a fiber section \(i\) of length \(L_i\), the Raman noise contributions are given by:
\begin{equation}
P_{\text{BS},i} = \frac{1 - e^{-(\alpha_s + \alpha_n)L_i}}{\alpha_s + \alpha_n} \beta_{\text{BS}}(\lambda_s, \lambda_n) \Delta\nu \, P_{\text{in}}
\end{equation}
\begin{equation}
P_{\text{FS},i} = \frac{e^{-\alpha_n L_i} - e^{-\alpha_s L_i}}{\alpha_s - \alpha_n} \beta_{\text{FS}}(\lambda_s, \lambda_n) \Delta\nu \, P_{\text{in}}
\end{equation}
where \(L_i\) is the length of fiber section \(i\), \(\alpha_s\) and \(\alpha_n\) are the attenuation constants at the classical signal wavelength \(\lambda_s\) and quantum channel wavelength \(\lambda_n\), respectively, \(P_{\text{in}}\) is the classical signal input power (in photons/s), \(\Delta\nu\) is the quantum channel bandwidth, and \(\beta_{\text{BS}}\) and \(\beta_{\text{FS}}\) are the backward and forward scattering conversion constants (in m\(^{-1}\)Hz\(^{-1}\)).

The scattering constants are wavelength-pair dependent and we use the experimentally measured values from \cite{burenkov2023synchronization}. For example, using 1270~nm classical signals with 1550~nm quantum channels yields the lowest noise (\(\beta_{\text{FS}} = 0.058 \times 10^{-23}\) m\(^{-1}\)Hz\(^{-1}\)), while 1490~nm classical signals produce approximately 64 times more noise.

For multi-section fibers, Raman noise is computed independently for each section where classical coexistence is enabled, and the total noise rate is obtained by summing all section contributions: $P_{\text{noise}} = \sum_i \left(P_{\text{FS},i} + P_{\text{BS},i}\right)$.

Raman noise photon detections are then generated stochastically following a Poisson process with rate \(P_{\text{noise}}\). Since spontaneous Raman scattering is a random process with constant average rate, the inter-arrival times between successive noise photon detections are sampled from an exponential distribution with mean \(1/P_{\text{noise}}\).

\section{Validation}
\label{sec:validation}
To validate our extensions to the SeQUeNCe simulator, we perform three levels of evaluation: component-level validation to verify individual optical elements, physical effect validation to confirm our fiber propagation models, and system-level validation through reproduction of published experimental results.

\subsection{Component-Level Validation}

\subsubsection{SPDC Source}
We validate our SPDC source implementation against experimental characterization of the Thorlabs SPDC810 source by Farella et al.~\cite{farella2024spectral}, which uses a periodically-poled KTP crystal with a 405~nm pump laser. Our simulation was configured with their measured parameters at three pump power levels (50, 100, 150~mW): signal and idler wavelengths near 810~nm and spectral bandwidths. The source operates at 80~MHz with mean photon number $\mu = 0.1$ per pulse, following thermal statistics.

\begin{figure}[h]
    \centering
    \includegraphics[width=\columnwidth]{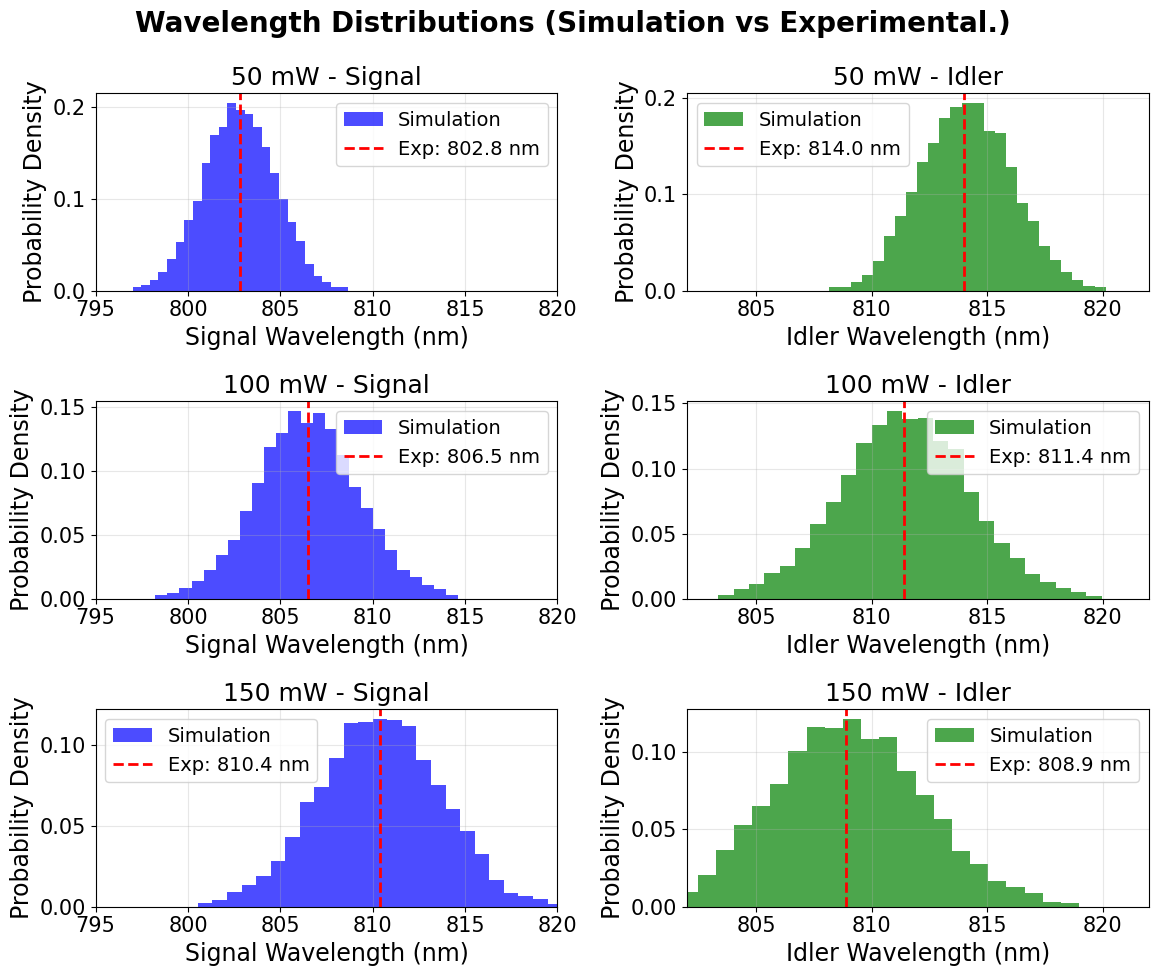}
    \caption{Wavelength distributions of signal (top) and idler (bottom) photons at 50, 100, and 150~mW pump powers. Histograms show simulation results; red dashed lines indicate experimental median values from Ref.~\cite{farella2024spectral}.}
    \label{fig:spdc_wavelength_dist}
\end{figure}

Figure~\ref{fig:spdc_wavelength_dist} compares simulated and experimental wavelength distributions. Signal photon distributions (top row, blue) closely match experimental median wavelengths (red dashed lines). Idler photon distributions (bottom row, green) emerge from energy conservation without direct parameterization, reproducing the experimental measurements.

\begin{figure*}[h]
    \centering
    \includegraphics[width=0.7\linewidth]{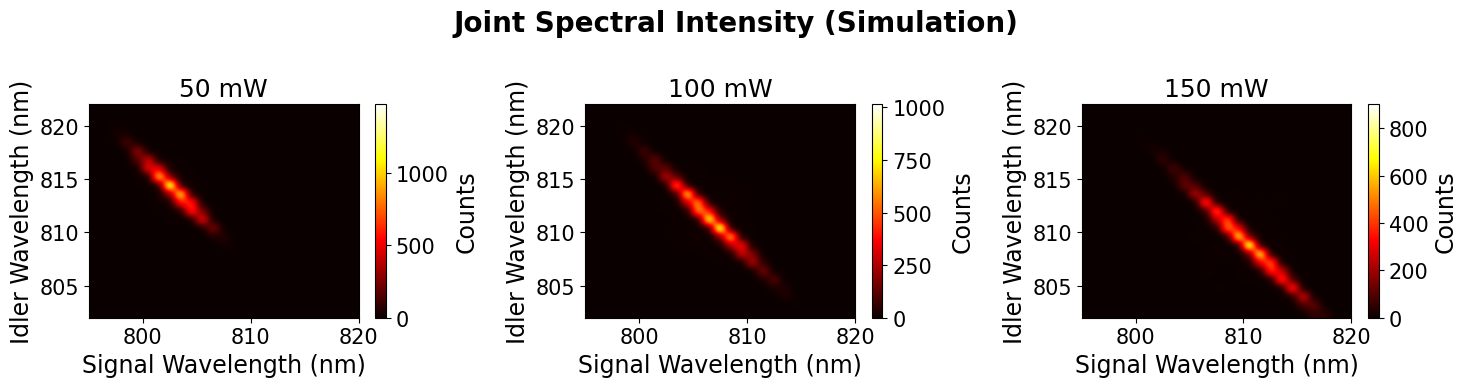}
    \caption{Joint Spectral Intensity showing signal-idler wavelength anti-correlation at 50~mW, 100~mW and 150~mW.}
    \label{fig:spdc_jsi}
\end{figure*}

Figure~\ref{fig:spdc_jsi} shows the Joint Spectral Intensity (JSI), which characterizes signal-idler wavelength correlations. The diagonal anti-correlation pattern (bright stripe) demonstrates energy conservation. The anti-correlation emerges naturally from enforcing energy conservation in the wavelength sampling procedure.

\subsubsection{Waveplate and Analyzer}

\begin{figure}[h]
    \centering
    \includegraphics[width=0.8\linewidth]{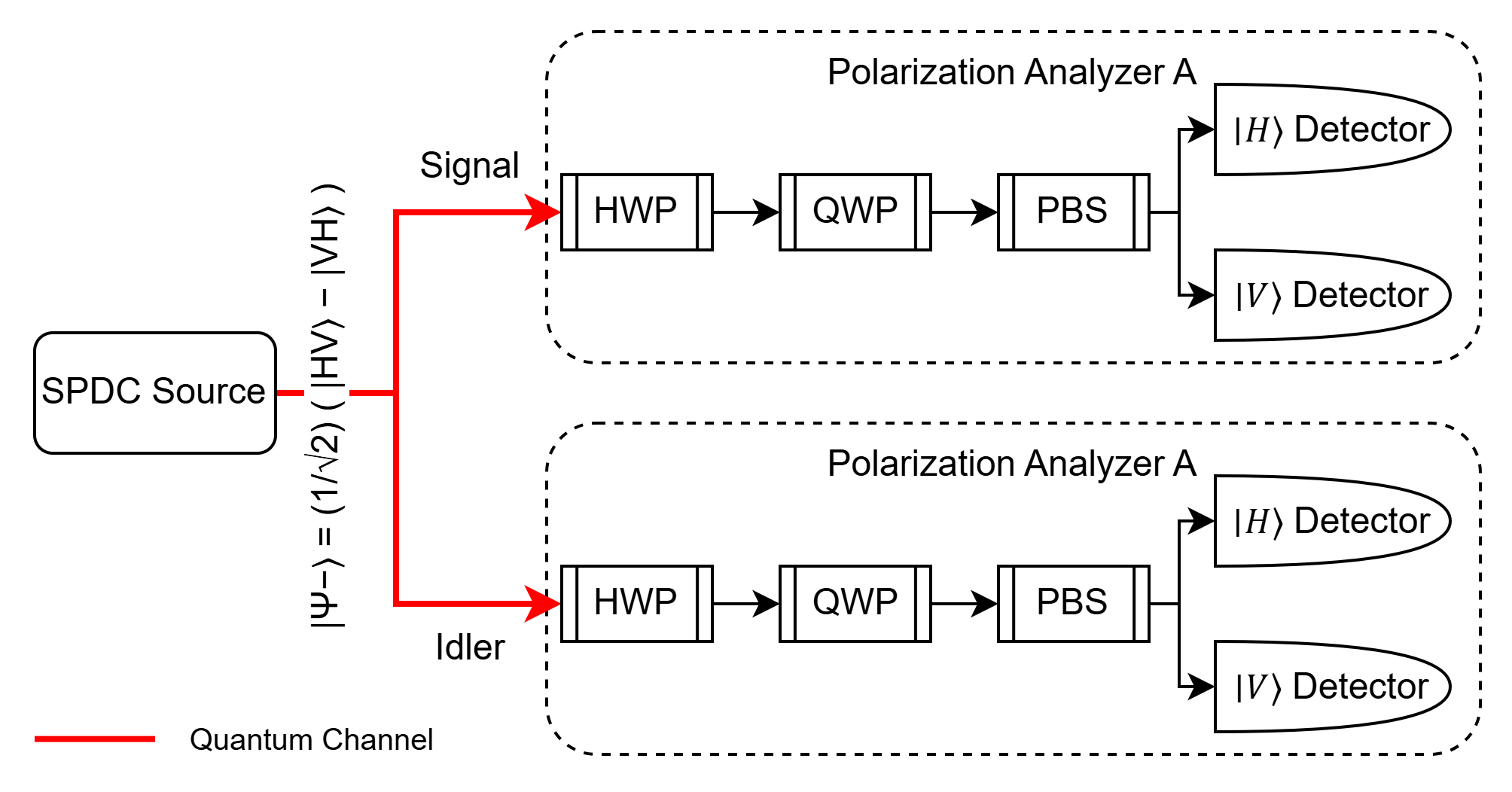}
    \caption{Simulation setup used for validating Waveplates and Analyzer}
    \label{fig:exp-setup}
\end{figure}

We validate the polarization measurement components by simulating a Bell state analyzer using the setup shown in Fig.~\ref{fig:exp-setup}. where an SPDC source emitting photon pairs in the state $|\Psi^-\rangle$. Each photon is directed to a polarization analyzer comprising a half-wave plate (HWP), quarter-wave plate (QWP), and polarizing beamsplitter (PBS) followed by two single-photon detectors. The HWP and QWP angles are configured to measure in a custom basis for correlation measurement, or different Pauli bases (Z, X, Y) for quantum state tomography. The simulation is performed without introducing any fiber or optical components losses.

\begin{figure}[h]
    \centering
    \includegraphics[width=0.8\columnwidth]{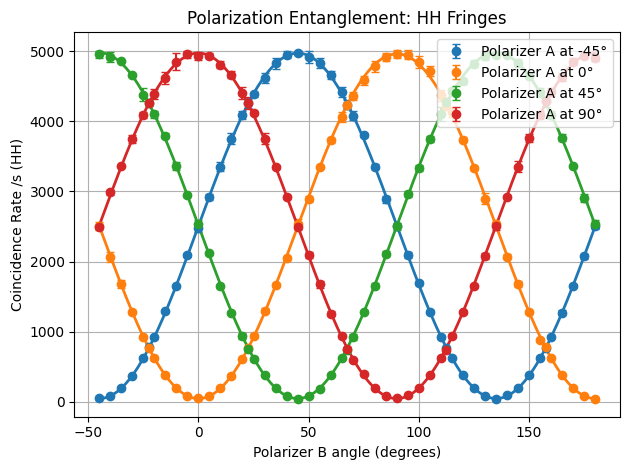}
    \caption{Polarization correlation fringes validating HWP functionality. Coincidence rates for Bell state $|\Psi^-\rangle$ measured at different polarizer settings.}
    \label{fig:hwp_validation}
\end{figure}

\paragraph{Correlation Measurements}
Figure~\ref{fig:hwp_validation} shows two-photon coincidence rates as polarizer B is scanned from $-45^\circ$ to $180^\circ$ with polarizer A held at four fixed angles ($-45^\circ$, $0^\circ$, $45^\circ$, $90^\circ$). The sinusoidal fringes with 180° period and 45° phase shifts between curves confirm correct HWP rotation. The data follows the expected $\cos^2(\theta_A - \theta_B)$ dependence for polarization-entangled photon pairs, with visibility exceeding 98\%.

\begin{figure}[ht]
    \centering
    \includegraphics[width=\columnwidth]{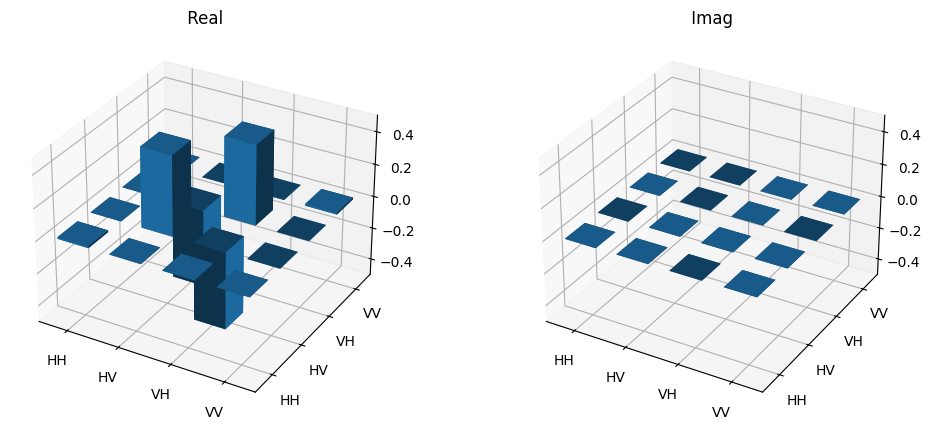}
    \caption{Reconstructed density matrix (real and imaginary parts) for $|\Psi^-\rangle$ from quantum state tomography.}
    \label{fig:tomography_validation}
\end{figure}

\paragraph{Quantum State Tomography}
Figure~\ref{fig:tomography_validation} shows the reconstructed density matrix from measurements in the Z, X, and Y bases on both analyzers. The measured density matrix in the computational basis $\{|HH\rangle, |HV\rangle, |VH\rangle, |VV\rangle\}$ is:

\[
\rho_{\text{measured}} \approx \begin{pmatrix}
0 & 0 & 0 & 0 \\
0 & 0.50 & -0.49 & 0 \\
0 & -0.49 & 0.50 & 0 \\
0 & 0 & 0 & 0
\end{pmatrix}
 \approx |\Psi^-\rangle\langle\Psi^-|
\]
\noindent where the entries displayed as $0$ are not exactly zero but satisfy $|\rho_{ij}| < 10^{-3}$.

The anti-diagonal structure with negative off-diagonal coherences $\rho_{HV,VH} = \rho_{VH,HV} \approx -0.49$ and near-zero populations in $|HH\rangle$ and $|VV\rangle$ confirm correct state preparation and measurement. State fidelity $F = \langle\Psi^-|\rho_{\text{measured}}|\Psi^-\rangle \approx 0.985$ validates the HWP+QWP tomography implementation.

\subsection{Physical Effect Validation}
\subsubsection{Polarization Mode Dispersion}
We validate the PMD model in the fiber by examining its impact on polarization correlations of entangled photon pairs. A continuous SPDC source emits photon pairs in the Bell state $\ket{\Psi^-} = (\ket{HV} - \ket{VH})/\sqrt{2}$. One photon is sent directly to a polarization analyzer, while the other is transmitted through a single-mode fiber in which we vary the mechanical stress. For simplicity we focus on fiber twist, but the same procedure can be applied to other perturbations such as bending or axial tension.

\begin{figure}[ht]
    \centering
    \includegraphics[width=\columnwidth]{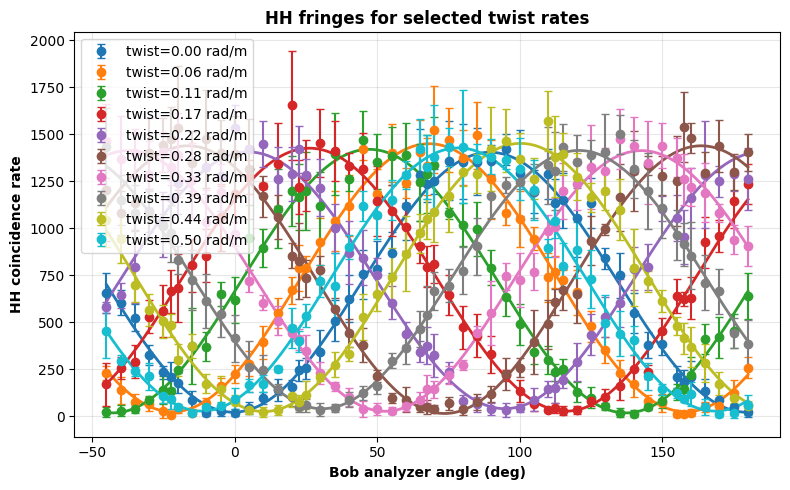}
    \caption{$HH$ coincidence rates versus Bob's analyzer angle for different fiber twist rates.}
    \label{fig:twist_fringes}
\end{figure}

Figure~\ref{fig:twist_fringes} shows the simulated $HH$ coincidence rate as a function of Bob’s analyzer angle for ten different fiber twist rates between $0$ and $0.5~\text{rad/m}$. Each curve is obtained by fixing Alice’s analyzer and scanning Bob’s half-wave plate, reproducing a standard polarization-correlation fringe measurement. As the twist rate increases, the fringes exhibit clear phase shifts, indicating that the fiber model primarily introduces unitary polarization rotations and PMD-induced phase changes.

The observed behavior matches the expected qualitative impact of PMD in telecom fibers: birefringence and mechanical stress change the effective principal states of polarization and their relative group delay, which appears as a controllable rotation of the correlation fringes that must be compensated in polarization-encoded quantum links, since any uncontrolled rotation directly translates into reduced polarization correlation and therefore lower fidelity of distributed entanglement.

\subsubsection{Chromatic Dispersion}
We next validate the chromatic dispersion model by analyzing coincidence timing for entangled photons transmitted through fibers of different lengths. A continuous SPDC source emits pairs in the Bell state $\ket{\Psi^-}$ with a finite spectral bandwidth around 1550~nm, so that different frequency components experience different delays. Each photon propagates through an identical single-mode fiber and is then measured in the H/V basis by polarization analyzers without additional losses.

\begin{figure*}[ht]
    \centering
    \includegraphics[width=\textwidth]{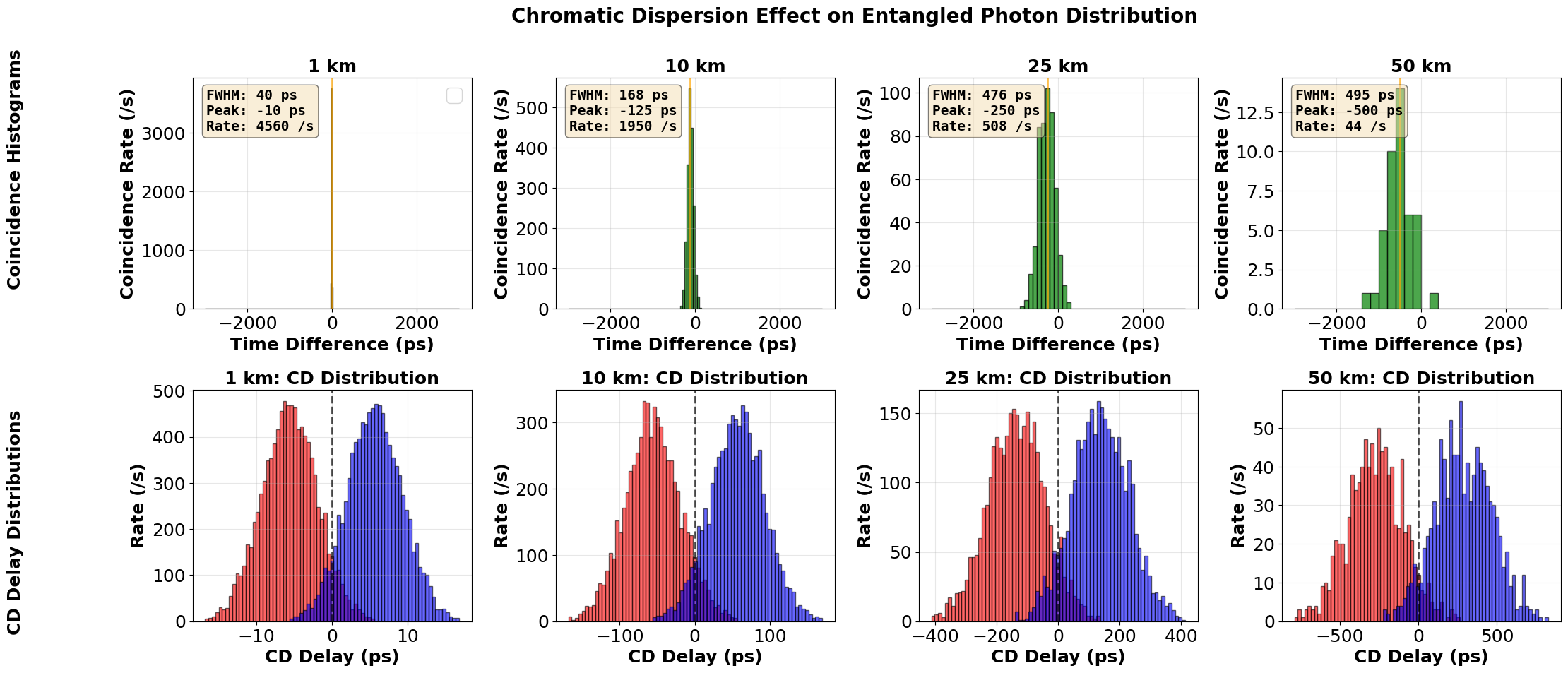}
    \caption{Effect of chromatic dispersion on entangled photon timing. Top: coincidence histograms of arrival-time difference for fiber lengths of 1, 10, 25, and 50~km. Bottom: chromatic dispersion delay distributions for the two channels at the same distances, extracted from the fiber model.}
    \label{fig:cd_analysis}
\end{figure*}

Figure~\ref{fig:cd_analysis} summarizes the results for fiber lengths of 1, 10, 25, and 50~km per arm. The top row shows coincidence histograms of the arrival-time difference between the two detectors in each arm. As the fiber length increases, the coincidence peak broadens and shifts, reflecting both the increased chromatic delay and the widening temporal spread induced by chromatic dispersion. For each distance we extract the full width at half maximum (FWHM) and peak position, which grow with length.

The bottom row shows the distributions of chromatic-dispersion-induced delays for channels A and B obtained directly from the fiber model. The delay histograms widen with distance and remain approximately symmetric around zero, confirming that both arms experience comparable dispersion. Together, the coincidence histograms and per-channel delay distributions demonstrate that the implemented fiber model produces physically consistent chromatic dispersion effects on entangled photon timing.

\subsubsection{Raman Scattering}
We validate our Raman noise implementation using the experimentally measured scattering coefficients reported by Burenkov et al.~\cite{burenkov2023synchronization}, which parameterize their analytical Raman-scattering model for spontaneous Raman noise in quantum channels coexisting with classical synchronization signals. Using this model, we reproduce the predicted Raman noise behavior across four classical wavelengths (1270~nm, 1310~nm, 1330~nm, and 1490~nm) with a quantum channel at 1550~nm and a 100~GHz detection bandwidth, demonstrating both distance and power dependence.

\begin{figure}[h]
\centering
\includegraphics[width=0.8\columnwidth]{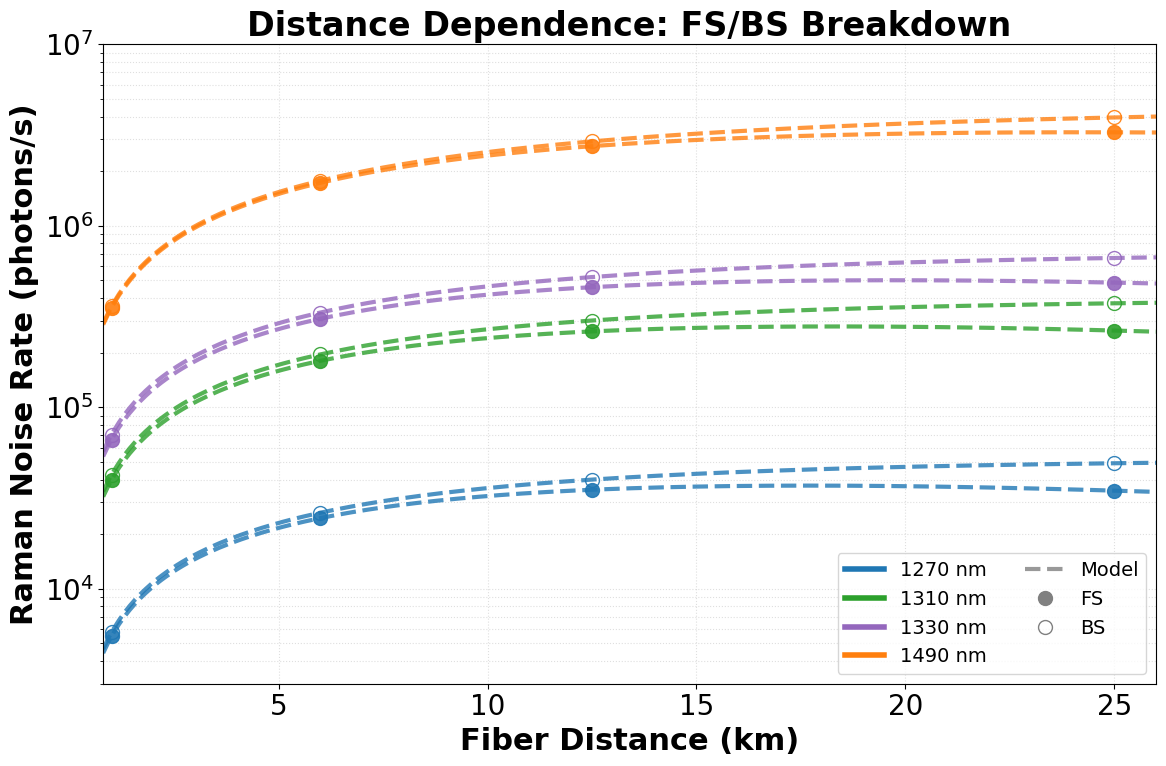}
\caption{Forward-scattering (FS) and backward-scattering (BS) Raman noise rate contributions versus fiber distance for four classical wavelengths, based on the analytical model using coefficients from Ref.~\cite{burenkov2023synchronization}.}
\label{fig:raman_distance_fsbs}
\end{figure}

\paragraph{Distance Dependence}
Figure~\ref{fig:raman_distance_fsbs} shows the forward-scattering (FS) and backward-scattering (BS) noise contributions as a function of fiber length for a fixed classical launch power of $10^{14}$~photons/s. The model curves illustrate how FS and BS components each contribute to the total Raman noise, with both increasing with distance and exhibiting characteristic saturation at longer fiber lengths. The 1490~nm classical signal produces approximately 64$\times$ more noise than the optimal 1270~nm wavelength, consistent with the scattering coefficients measured by Burenkov et al.~\cite{burenkov2023synchronization}.

\begin{figure}[h]
\centering
\includegraphics[width=0.8\columnwidth]{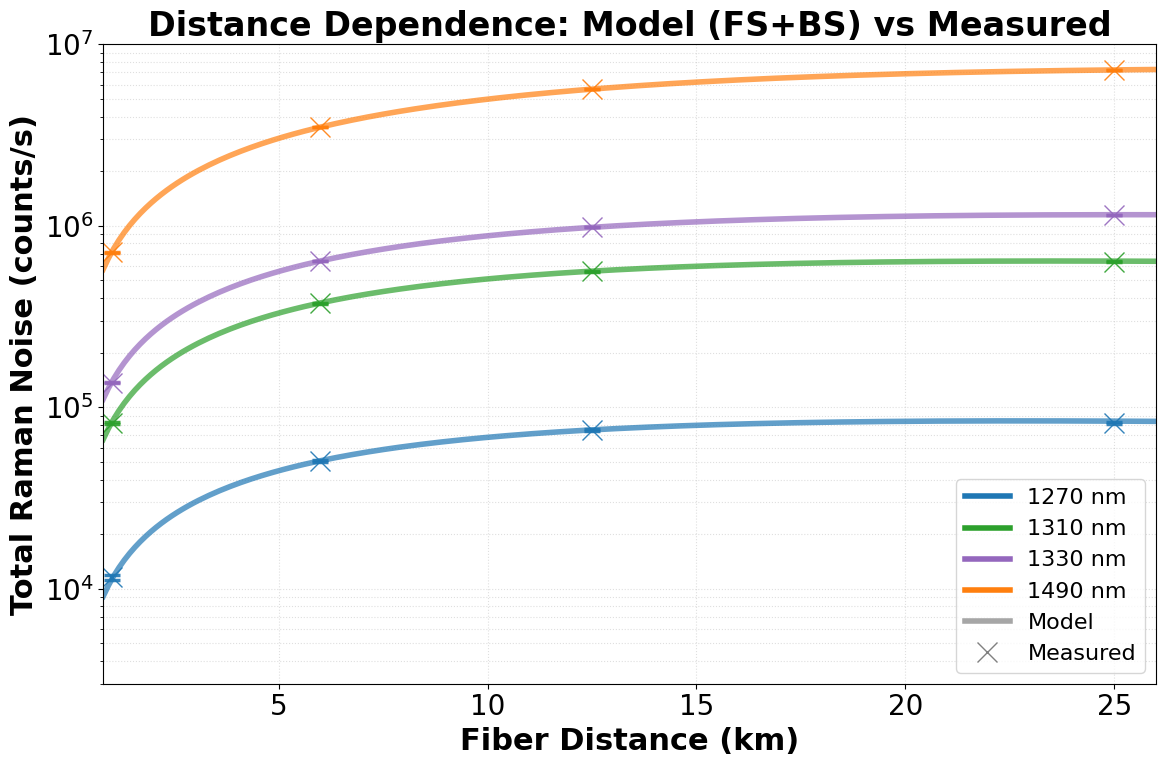}
\caption{Total Raman noise rate (FS+BS) versus fiber distance: analytical model predictions from Ref.~\cite{burenkov2023synchronization} (solid lines) compared to simulation measurements (X markers).}
\label{fig:raman_distance_validation}
\end{figure}

Figure~\ref{fig:raman_distance_validation} compares the total Raman noise rate (FS+BS) predicted by the analytical model against simulation measurements. The model curves (solid lines) show excellent agreement with the simulation data (X markers) across all wavelengths and distances from 1~km to 25~km. The total noise increases monotonically with fiber length, with the growth rate gradually decreasing at longer distances due to saturation effects. This behavior matches the theoretical predictions and real measurement data from Ref.~\cite{burenkov2023synchronization}.

\paragraph{Power Dependence}

\begin{figure}[h]
\centering
\includegraphics[width=0.8\columnwidth]{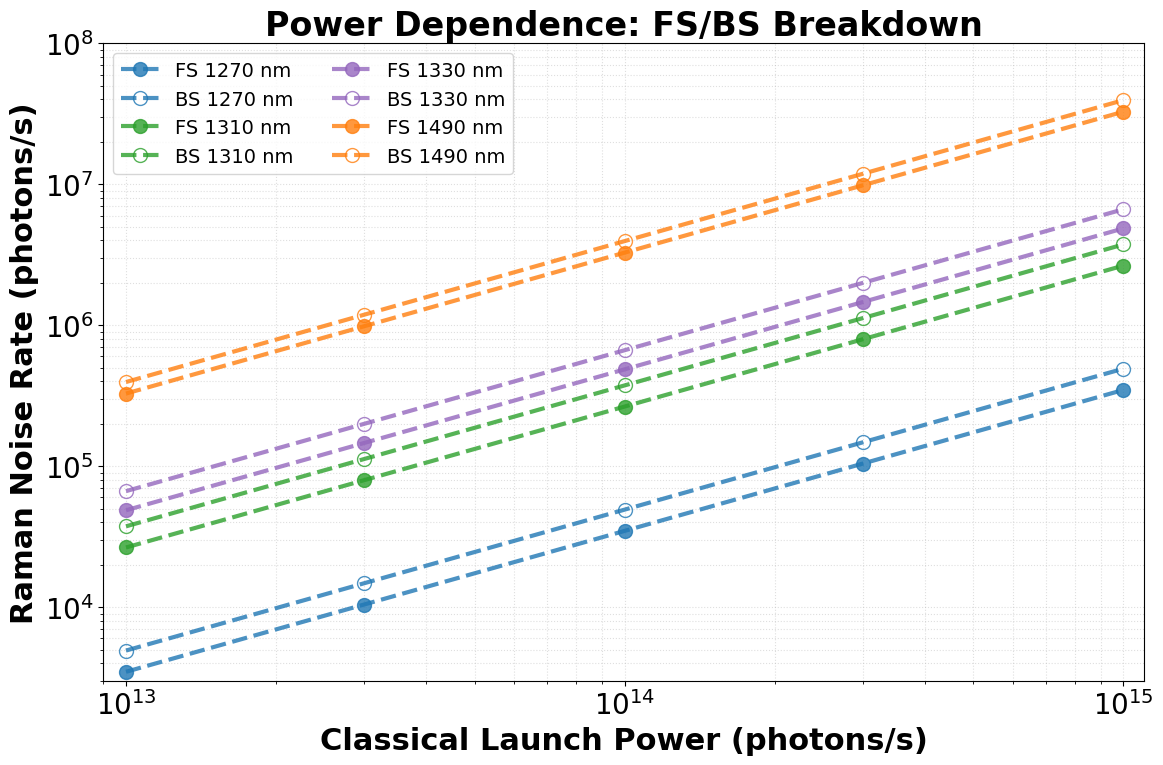}
\caption{Forward-scattering (FS) and backward-scattering (BS) Raman noise rate contributions versus classical launch power for four classical wavelengths.}
\label{fig:raman_power_fsbs}
\end{figure}

Figure~\ref{fig:raman_power_fsbs} illustrates the FS and BS noise contributions as a function of classical launch power for a fixed 25~km fiber length. The log-log representation shows the expected linear relationship between classical power and Raman noise rate: the noise scales proportionally with launched power across two orders of magnitude. FS and BS contributions are comparable in magnitude for all wavelengths, with both following the same power-law scaling.

\begin{figure}[h]
\centering
\includegraphics[width=0.8\columnwidth]{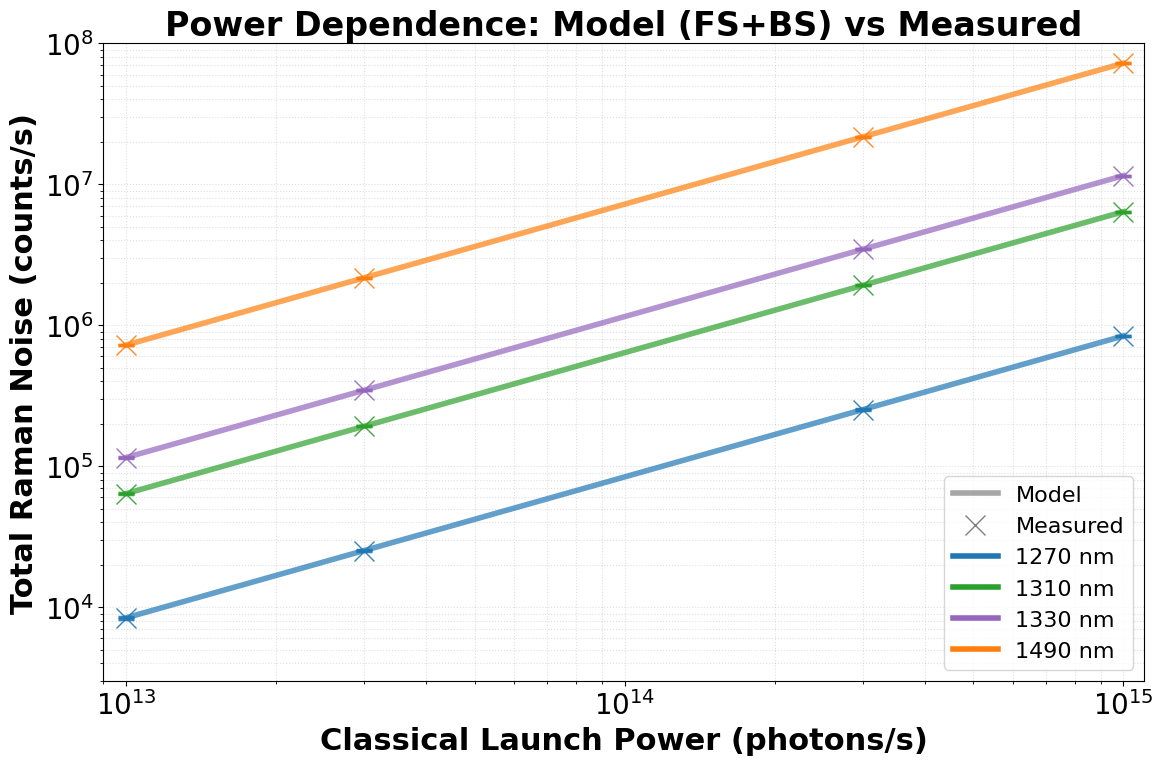}
\caption{Total Raman noise (FS+BS) versus classical launch power: analytical model predictions (solid lines); simulation measurements (X markers).}
\label{fig:raman_power_validation}
\end{figure}

Figure~\ref{fig:raman_power_validation} compares the analytical model predictions against simulation measurements across classical powers ranging from $10^{13}$ to $10^{15}$~photons/s. The model curves (solid lines) show strong agreement with simulation data (X markers) over this $100\times$ dynamic range for all four wavelengths. The log-log representation demonstrates the linear power scaling: Raman noise increases proportionally with classical signal power. The 1270~nm channel maintains the lowest noise, while 1490~nm consistently produces the highest noise, consistent with the measurements by Burenkov et al.~\cite{burenkov2023synchronization}.

\section{Discussion}
\label{sec:discussion}
Our framework focuses on polarization-encoded photonic qubits, reflecting current metropolitan deployments. However, other encodings such as time-bin or frequency-bin exhibit different sensitivities to fiber impairments. Extending the simulator to support multiple encodings would enable systematic comparison of physical-layer robustness across platforms.

The inclusion of Jones-matrix propagation and stochastic birefringence modeling increases computational overhead relative to abstract depolarization channels. To maintain scalability, we employ a segment-based fiber representation with event-driven updates rather than continuous-time integration. While suitable for metropolitan-scale simulations, very large networks or long-term drift studies may require hierarchical abstractions or multi-resolution techniques, a limitation common to detailed modeling of other quantum subsystems such as memories or circuits.

Although we validate our work through individual physical effects, we do not yet provide a comprehensive cross-layer evaluation of their impact on routing, swapping, or purification strategies. The framework enables such investigations, but systematic protocol-level studies remain future work.

Finally, predictive accuracy depends on deployment-specific parameterization. While grounded in experimentally reported constants, real fiber links exhibit site-dependent variability. This work should therefore be viewed as a physics-informed modeling platform that supports hardware-aware design exploration rather than a universally predictive tool. Our approach complements existing abstraction-oriented simulators by emphasizing cross-layer interpretability and direct parameterization by measurable optical properties.

Despite these limitations, embedding optical-layer physics into discrete-event quantum network simulation represents an essential step toward realistic quantum network engineering. By explicitly connecting fiber-level physical parameters to network-level performance metrics, the framework enables systematic investigation of classical–quantum coexistence strategies, hardware-informed protocol design, and future cross-layer optimization mechanisms such as adaptive calibration and tomography-driven link certification. Bridging the abstraction gap between optical physics and network protocols is critical for the transition from laboratory demonstrations to robust, deployable quantum communication infrastructures.

\section{Conclusion}
\label{sec:conclusion}
We have extended the SeQUeNCe discrete-event simulator with physics-based models for polarization-encoded, fiber-based quantum networks. By integrating Jones-calculus optical components and a parameterized fiber channel model capturing polarization mode dispersion, chromatic dispersion, and Raman noise, we bridge the gap between optical-layer physics and network-level simulation.

This framework enables hardware-informed evaluation of entanglement distribution under realistic deployment conditions and provides a foundation for cross-layer studies of protocol performance in practical quantum networks.


\bibliographystyle{IEEEtran}  
\bibliography{references}

@article{wengerowsky2019entanglement,
  title={Entanglement distribution over a 96-km-long submarine optical fiber},
  author={Wengerowsky, S{\"o}ren and Joshi, Siddarth Koduru and Steinlechner, Fabian and Zichi, Julien R and Dobrovolskiy, Sergiy M and Van der Molen, Rene and Los, Johannes WN and Zwiller, Val and Versteegh, Marijn AM and Mura, Alberto and others},
  journal={Proceedings of the National Academy of Sciences},
  volume={116},
  number={14},
  pages={6684--6688},
  year={2019},
  publisher={National Academy of Sciences}
}

@article{zhang2008distribution,
  title={Distribution of time-energy entanglement over 100 km fiber using superconducting single-photon detectors},
  author={Zhang, Qiang and Takesue, Hiroki and Nam, Sae Woo and Langrock, Carsten and Xie, Xiuping and Baek, Burm and Fejer, Martin M and Yamamoto, Yoshihisa},
  journal={Optics express},
  volume={16},
  number={8},
  pages={5776--5781},
  year={2008},
  publisher={Optical Society of America}
}

@article{ikuta2018four,
  title={Four-dimensional entanglement distribution over 100 km},
  author={Ikuta, Takuya and Takesue, Hiroki},
  journal={Scientific reports},
  volume={8},
  number={1},
  pages={817},
  year={2018},
  publisher={Nature Publishing Group UK London}
}

@article{rahmouni2024100,
  title={100-km entanglement distribution with coexisting quantum and classical signals in a single fiber},
  author={Rahmouni, Anouar and Kuo, PS and Li-Baboud, Ya-Shian and Burenkov, IA and Shi, Yicheng and Jabir, MV and Lal, N and Reddy, Dileep and Merzouki, Mheni and Ma, Lijun and others},
  journal={Journal of Optical Communications and Networking},
  volume={16},
  number={8},
  pages={781--787},
  year={2024},
  publisher={Optica Publishing Group}
}

@article{burenkov2023synchronization,
  title={Synchronization and coexistence in quantum networks},
  author={Burenkov, Ivan A and Semionov, Alexandra and Hala and Gerrits, Thomas and Rahmouni, Anouar and Anand, DJ and Li-Baboud, Ya-Shian and Slattery, Oliver and Battou, Abdella and Polyakov, Sergey V},
  journal={Optics Express},
  volume={31},
  number={7},
  pages={11431--11446},
  year={2023},
  publisher={Optica Publishing Group}
}

@article{bennett1992quantum,
  title={Quantum cryptography using any two nonorthogonal states},
  author={Bennett, Charles H},
  journal={Physical review letters},
  volume={68},
  number={21},
  pages={3121},
  year={1992},
  publisher={APS}
}

@article{ekert1991quantum,
  title={Quantum cryptography based on Bell’s theorem},
  author={Ekert, Artur K},
  journal={Physical review letters},
  volume={67},
  number={6},
  pages={661},
  year={1991},
  publisher={APS}
}

@article{hubel2007high,
  title={High-fidelity transmission of polarization encoded qubits from an entangled source over 100 km of fiber},
  author={H{\"u}bel, Hannes and Vanner, Michael R and Lederer, Thomas and Blauensteiner, Bibiane and Lor{\"u}nser, Thomas and Poppe, Andreas and Zeilinger, Anton},
  journal={Optics Express},
  volume={15},
  number={12},
  pages={7853--7862},
  year={2007},
  publisher={Optical Society of America}
}

@article{jones1941new,
  title={A new calculus for the treatment of optical systemsi. description and discussion of the calculus},
  author={Jones, R Clark},
  journal={Journal of the Optical Society of America},
  volume={31},
  number={7},
  pages={488--493},
  year={1941},
  publisher={OSA}
}

@article{le1997utilization,
  title={Utilization of Mueller matrix formalism to obtain optical targets depolarization and polarization properties},
  author={Le Roy-Brehonnet, F and Le Jeune, B},
  journal={Progress in Quantum Electronics},
  volume={21},
  number={2},
  pages={109--151},
  year={1997},
  publisher={Elsevier}
}

@article{gordon2000pmd,
  title={PMD fundamentals: Polarization mode dispersion in optical fibers},
  author={Gordon, JP and Kogelnik, H},
  journal={Proceedings of the National Academy of Sciences},
  volume={97},
  number={9},
  pages={4541--4550},
  year={2000},
  publisher={The National Academy of Sciences}
}

@incollection{agrawal2000nonlinear,
  title={Nonlinear fiber optics},
  author={Agrawal, Govind P},
  booktitle={Nonlinear Science at the Dawn of the 21st Century},
  pages={195--211},
  year={2000},
  publisher={Springer}
}

@article{kwiat1995new,
  title={New high-intensity source of polarization-entangled photon pairs},
  author={Kwiat, Paul G and Mattle, Klaus and Weinfurter, Harald and Zeilinger, Anton and Sergienko, Alexander V and Shih, Yanhua},
  journal={Physical Review Letters},
  volume={75},
  number={24},
  pages={4337},
  year={1995},
  publisher={APS}
}

@article{carvalho2025spectrally,
  title={Spectrally Resolved Higher Order Photon Statistics of Spontaneous Parametric Down Conversion},
  author={Carvalho, Jeffrey and Wijesundara, Chiran and Thomay, Tim},
  journal={arXiv preprint arXiv:2505.22883},
  year={2025}
}

@misc{thorlabs_pbs,
  title={Broadband Polarizing Beamsplitter Cubes},
  author={{Thorlabs Inc.}},
  howpublished={\url{https://www.thorlabs.com}},
  note={Accessed: 2026}
}

@article{farella2024spectral,
  title={Spectral characterization of an SPDC source with a fast broadband spectrometer},
  author={Farella, Brianna and Medwig, Gregory and Abrahao, Raphael A and Nomerotski, Andrei},
  journal={AIP Advances},
  volume={14},
  number={4},
  year={2024},
  publisher={AIP Publishing}
}

@article{wu2021sequence,
  title={SeQUeNCe: a customizable discrete-event simulator of quantum networks},
  author={Wu, Xiaoliang and Kolar, Alexander and Chung, Joaquin and Jin, Dong and Zhong, Tian and Kettimuthu, Rajkumar and Suchara, Martin},
  journal={Quantum Science and Technology},
  volume={6},
  number={4},
  pages={045027},
  year={2021},
  publisher={IOP Publishing}
}

@article{abane2025multiverse,
  title={Multiverse: A Simulator for Evaluating Entanglement Routing in Quantum Networks},
  author={Abane, Amar and Shi, Junxiao and Mai, Van Sy and Amlou, Abderrahim and Battou, Abdella},
  journal={arXiv preprint arXiv:2512.22937},
  year={2025}
}

@article{coopmans2021netsquid,
  title={Netsquid, a network simulator for quantum information using discrete events},
  author={Coopmans, Tim and Knegjens, Robert and Dahlberg, Axel and Maier, David and Nijsten, Loek and de Oliveira Filho, Julio and Papendrecht, Martijn and Rabbie, Julian and Rozpedek, Filip and Skrzypczyk, Matthew and others},
  journal={Communications Physics},
  volume={4},
  number={1},
  pages={164},
  year={2021},
  publisher={Nature Publishing Group UK London}
}

@article{james2001measurement,
  title={Measurement of qubits},
  author={James, Daniel FV and Kwiat, Paul G and Munro, William J and White, Andrew G},
  journal={Physical Review A},
  volume={64},
  number={5},
  pages={052312},
  year={2001},
  publisher={APS}
}

@article{banner2025bifrost,
  title={BIFROST: A First-Principles Model of Polarization Mode Dispersion in Optical Fiber},
  author={Banner, Patrick R and Rolston, Steven L and Britton, Joseph W},
  journal={arXiv preprint arXiv:2510.01212},
  year={2025}
}

@inproceedings{amlou2025scalable,
  title={Scalable Time-Tagged Data Acquisition for Entanglement Distribution in Quantum Networks},
  author={Amlou, Abderrahim and Gerrits, Thomas and Rahmouni, Anouar and Abane, Amar and Merzouki, Mheni and Li-Baboud, Ya-Shian and Lbath, Ahmed and Battou, Abdella and Slattery, Oliver},
  booktitle={2025 IEEE International Conference on Quantum Computing and Engineering (QCE)},
  volume={1},
  pages={1134--1140},
  year={2025},
  organization={IEEE}
}
\end{document}